\documentclass[aps,prd,reprint,superscriptaddress,nofootinbib,longbibliography,floatfix]{revtex4-2}
\usepackage[T1]{fontenc}
\usepackage{amsmath,amssymb,bm,graphicx,booktabs,xcolor,capt-of,placeins}
\usepackage[caption=false,font=footnotesize,subrefformat=parens]{subfig}
\usepackage{tikz}
\usepackage{needspace}
\usetikzlibrary{calc}
\newcommand{\plotpanelabove}[4]{%
 \subfloat{\label{#2}%
  \begin{tikzpicture}[baseline=(p.south),inner sep=0pt,outer sep=0pt]
   \node[anchor=south west,inner sep=0pt,outer sep=0pt] (p) at (0,0) {\includegraphics[width=#3]{#1}};
   \begin{scope}[shift={(p.south west)},x={($(p.south east)-(p.south west)$)},y={($(p.north west)-(p.south west)$)}]
    \node[anchor=south west,inner sep=0pt,outer sep=0pt] at (#4) {\normalsize\bfseries(\thesubfigure)};
   \end{scope}
  \end{tikzpicture}}}

\usepackage[colorlinks=true,linkcolor=blue,citecolor=blue,urlcolor=blue]{hyperref}
\hypersetup{pdftitle={Boundary preparation of strong Mpemba states in holography},pdfauthor={Yu Shi, Xinwu Liu, Liping Li, Chikun Ding, Yuebing Zhou, Changqing Liu, Sheng Long}}
\graphicspath{{figures/}}
\AddToHook{env/figure/begin}{\setlength{\abovecaptionskip}{9pt}}
\AddToHook{env/figure*/begin}{\setlength{\abovecaptionskip}{9pt}}

\newcommand{\ii}{\mathrm{i}}
\newcommand{\dd}{\mathrm{d}}
\newcommand{\cB}{\mathcal B}
\newcommand{\cN}{\mathcal N}
\newcommand{\cI}{\mathcal I}
\newcommand{\cC}{\mathcal C}
\newcommand{\cO}{\mathcal O}
\newcommand{\Real}{\operatorname{Re}}
\newcommand{\Imag}{\operatorname{Im}}
\newcommand{\ord}[1]{\mathcal O(#1)}
\newif\ifmarkrevisions
\markrevisionsfalse

\begin{document}

\title{Boundary preparation of strong Mpemba states in holography}

\author{Yu Shi}
\email{shiyu@hhtc.edu.cn}
\affiliation{Department of Physics, Huaihua University, Huaihua, Hunan 418008, China}

\author{Xinwu Liu}
\affiliation{Department of Physics, Huaihua University, Huaihua, Hunan 418008, China}

\author{Liping Li}
\affiliation{Department of Physics, Huaihua University, Huaihua, Hunan 418008, China}

\author{Chikun Ding}
\affiliation{Department of Physics, Huaihua University, Huaihua, Hunan 418008, China}

\author{Yuebing Zhou}
\affiliation{Department of Physics, Huaihua University, Huaihua, Hunan 418008, China}

\author{Changqing Liu}
\affiliation{Department of Physics, Huaihua University, Huaihua, Hunan 418008, China}

\author{Sheng Long}
\email[Corresponding author: ]{shenglong@ucas.ac.cn}
\affiliation{School of Fundamental Physics and Mathematical Sciences, Hangzhou Institute for Advanced Study, University of Chinese Academy of Sciences, Hangzhou 310024, China}

\begin{abstract}
We study the boundary preparation of strong Mpemba states in a probe Abelian Higgs model on a fixed AdS black brane. Different drives ending at the same charge density can prepare states with different slow mode amplitudes and relaxation behavior. Two pulses cancel the slowest mode at linear order, but charge exchange during preparation and radial Maxwell feedback regenerate the slow tail at cubic order. We derive the complete cubic correction from the lower order field equations, including feedback after source removal, and use it to predict a compensating pulse. Full nonlinear evolution confirms that cubic compensation reduces the residual slow amplitude from $O(\epsilon^3)$ to $O(\epsilon^5)$. For a representative protocol at final density $\rho_f=3$ and $\epsilon=0.05$, cubic compensation suppresses the slow amplitude by a factor of about $600$ relative to the linear prescription. Cubic compensation extends the period of fast relaxation without changing the final quasinormal mode spectrum. Further nonlinear tuning suppresses the residual slow amplitude to the level of numerical resolution. The excess free energy of the prepared states can exceed that of a weakly driven reference at release and fall below it during subsequent relaxation. Three pulses also suppress a second slow channel over the resolved time window and reveal interference between faster modes. Starting from a common initial state, controls with zero net charge transfer reach a given relaxation threshold sooner than passive relaxation, even when the full control duration is included.
\end{abstract}

\maketitle
\raggedbottom

\section{Introduction}
\label{sec:introduction}

A hotter system is generally expected to take longer to cool than a colder one in the same environment. The Mpemba effect challenges this expectation. Under suitable conditions, a system that is initially hotter can approach equilibrium faster~\cite{Mpemba1969,Teza2026Review}. This counterintuitive behavior has been observed in colloidal experiments, both in cooling and in the inverse process of faster heating~\cite{Kumar2020Cooling,Kumar2022Heating}. Near equilibrium, a useful dynamical perspective comes from the spectrum of relaxation modes. The evolution depends on both the decay rates of these modes and the amplitudes with which they are excited~\cite{LuRaz2017}. Since faster modes decay first, the slowest mode generally dominates at late times, even if its initial amplitude is small. If preparation eliminates this leading slow contribution, faster modes govern the subsequent relaxation. A suitably prepared state can then relax faster than a state that starts closer to equilibrium. This gives rise to the strong Mpemba effect~\cite{Klich2019,BlomEtAl2026Reentrant}. Studies of Mpemba phenomena have since extended to symmetry restoration~\cite{Ares2023,AresCalabreseMurciano2025Review,Rylands2024,Murciano2024XY}, active mode suppression~\cite{Carollo2021}, multireservoir systems~\cite{NavaEgger2024}, experiments with trapped ions~\cite{Zhang2025,AharonyShapira2024,Joshi2024}, and relativistic quantum relaxation~\cite{Liu2026RelativisticStrong}.

Most discussions based on this spectral picture start from a prescribed initial state. In practice, however, preparing a state takes time. The system continues to evolve under the external drive during this period, so the preparation itself can influence its subsequent relaxation. Simulations of colloidal systems show that quenches at a finite rate can alter the Mpemba effect~\cite{SaneBechhoefer2026}. Studies of optimal control in granular fluids also show that preparation dynamics can restrict the accessible nonequilibrium states~\cite{RiosMonje2026}. These results show how preparation affects the states that can be reached and their subsequent relaxation.

Holography provides a useful framework for studying nonequilibrium dynamics in strongly coupled systems~\cite{Maldacena1998,GKP1998,Witten1998,LiuSonner2020}. In this framework, boundary sources drive the evolution of bulk fields, while quasinormal modes (QNMs) describe the linear decay toward equilibrium~\cite{HorowitzHubeny2000,KovtunStarinets2005}. Energy flux into the black hole horizon captures dissipation during relaxation~\cite{Tian2023}. Previous studies of holographic dynamics have addressed condensation, periodic driving, and relaxation near criticality~\cite{MurataKinoshitaTanahashi2010,Bhaseen2013,LiTianZhang2013,IshiiMurata2018,Flory2024}. More recently, Ref.~\cite{Ge2026} studied the nonlinear Mpemba effect in a holographic superfluid and related the anomalous relaxation to competition among the slowest QNMs. Its setup is equivalent to an instantaneous density quench, followed by relaxation at fixed density with no external source.

To study preparation by finite boundary driving, we use a homogeneous probe Abelian Higgs model on a fixed AdS black brane~\cite{HHH2008,HHH2008b}. The fixed geometry provides an unchanged thermal background, while the coupled scalar and Maxwell fields continue to evolve. During the drive, the boundary source exchanges charge with the system and changes the radial electric field, affecting subsequent scalar relaxation. We match the final charge density of different preparations so that they approach the same normal equilibrium. This allows us to examine how preparation affects subsequent relaxation while the linear decay spectrum remains the same.

We focus on how finite preparation determines the slow response during subsequent relaxation. During preparation, charge exchange and Maxwell feedback contribute to the slow response. Maxwell feedback continues after the source is removed. The complete asymptotic slow amplitude therefore contains contributions from both stages, and cancellation at linear order generally does not eliminate it. The cubic structure of the feedback allows us to calculate the leading nonlinear correction to this amplitude from the field equations at lower orders and predict a compensating pulse. We then use the full nonlinear evolution to test this prediction and its effect on relaxation.

We further examine how a preparation with three pulses controls two slow channels within the time window studied. We also study charge neutral protocols starting from a common input state. The rest of the paper is organized as follows. Section~\ref{sec:model} introduces the model and the relaxation observables. Section~\ref{sec:cancellation} describes the nonlinear slow response and its cancellation. Section~\ref{sec:preparation} presents preparation with two pulses and the fast relaxation window. Section~\ref{sec:two-mode-preparation} studies preparation with three pulses and interference between fast modes. Section~\ref{sec:common} discusses charge neutral control from a common input. Section~\ref{sec:discussion} concludes. The Appendix provides details of the nonlinear mode amplitudes, pulse corrections, and numerical methods.

\section{Holographic model and relaxation observables}
\label{sec:model}

\subsection{Holographic model and boundary conditions}

We use the standard Abelian Higgs model of holographic superconductivity~\cite{HHH2008,HHH2008b}. The bulk theory contains a Maxwell field and a charged complex scalar coupled to Einstein gravity in four dimensions. The Maxwell field describes the boundary charge current, while the scalar describes the superconducting order parameter. The action is
\begin{equation}
 S=\frac{1}{16\pi G_4}\int\dd^4x\sqrt{-g}\left[R-2\Lambda+\mathcal L_{\mathrm m}\right],
 \label{eq:action}
\end{equation}
where $R$ is the Ricci scalar and $\Lambda=-3/L^2$, with $L$ the AdS radius. The matter Lagrangian is
\begin{equation*}
 \mathcal L_{\mathrm m}=-\tfrac14F_{ab}F^{ab}
 -(D_a\Phi)^*D^a\Phi-m^2|\Phi|^2,
\end{equation*}
with $F_{ab}=\partial_aA_b-\partial_bA_a$ and $D_a=\nabla_a-\ii qA_a$. The parameters $m$ and $q$ are the mass and charge of the scalar field.

We take the probe limit to study coupled scalar and charge dynamics on a fixed thermal background. The scalar and Maxwell fields remain coupled, while their gravitational backreaction is neglected. We rescale the fields as $A_a\to qA_a$ and $\Phi\to q\Phi$, with the symbols on the right denoting the fields used below. The matter action acquires a factor $q^{-2}$, which suppresses backreaction as $q\to\infty$ at fixed rescaled fields. The covariant derivative becomes $D_a=\nabla_a-\ii A_a$. We suppress the common overall normalization of the matter observables.

We choose a planar Schwarzschild AdS black brane as the fixed background. To describe the driven evolution and subsequent relaxation, we use ingoing Eddington Finkelstein coordinates, which are regular at the future horizon~\cite{LiTianZhang2013}. The metric is
\begin{equation}
 \dd s^2=\frac{L^2}{z^2}\left[-f(z)\dd t^2-2\dd t\dd z+\dd x^2+\dd y^2\right],
 \label{eq:metric}
\end{equation}
where
\begin{equation*}
 f(z)=1-\left(\frac{z}{z_h}\right)^3,\qquad T=\frac{3}{4\pi z_h}.
\end{equation*}
The conformal boundary is at $z=0$ and the future horizon is at $z=z_h$. We choose $m^2L^2=-2$ and use units in which $L=z_h=1$.

Varying the matter action gives the scalar and Maxwell equations in the rescaled variables,
\begin{align*}
 &(D_aD^a-m^2)\Phi=0,\\
 &\nabla_aF^{ab}-\ii\left[\Phi^*D^b\Phi-\Phi(D^b\Phi)^*\right]=0.
\end{align*}

We consider spatially uniform boundary drives, so the fields depend only on $t$ and $z$. In radial gauge, we write
\begin{equation*}
 \Phi=z\psi(t,z),\qquad A=A_t(t,z)\dd t.
\end{equation*}
We impose regularity at the future horizon and fix the remaining gauge freedom by setting $A_t(t,1)=0$. The scalar equation and radial Maxwell constraint then become
\begin{align}
 f\psi''-2\dot\psi'+(f'+2\ii A_t)\psi'
 +(\ii A_t'-z)\psi&=0,\label{eq:scalar}\\
 A_t''-\ii(\bar\psi\psi'-\psi\bar\psi')&=0.\label{eq:gauss}
\end{align}
Dots and primes denote derivatives with respect to $t$ and $z$. The scalar equation governs the field evolution, while the Maxwell constraint determines the radial electric potential from the scalar current and the boundary charge density.

The boundary expansion identifies the source and the response of the dual operator,
\begin{align*}
 \Phi(t,z)&=zJ(t)+z^2B(t)+O(z^3),\\
 A_t(t,z)&=\mu(t)-\rho(t)z+O(z^2).
\end{align*}
In standard quantization, $J$ is the source for a boundary operator of dimension two. The coefficients $\mu$ and $\rho$ are the boundary electric potential and charge density. In the chosen coordinates and gauge, holographic renormalization gives the expectation value~\cite{LiTianZhang2013}
\begin{equation}
 \langle\cO\rangle=B-\dot J+\ii\mu J.
 \label{eq:response}
\end{equation}
The scalar source carries charge, so it can exchange charge with the system during preparation. The charge density therefore evolves according to the Ward identity~\cite{Bianchi2002}
\begin{equation}
 \dot\rho=2\Imag\!\left(\bar J\langle\cO\rangle\right).
 \label{eq:ward}
\end{equation}
We use pulses of finite duration and denote the final source removal time by $t_0$. After the source is switched off, these relations reduce to
\begin{equation*}
 \langle\cO\rangle=B=\psi'(t,0),\qquad \dot\rho=0.
\end{equation*}
After source removal, the charge density remains fixed at $\rho_f$, while the scalar and electric profiles left by the drive continue to evolve.

We compare preparations that end at the same charge density $\rho_f$ and relax toward the same normal equilibrium,
\begin{equation}
 \psi_{\mathrm{eq}}=0,\qquad A_{t,\mathrm{eq}}=\rho_f(1-z).
 \label{eq:normal}
\end{equation}
The main pulse comparisons use $\rho_f=\rho_*=3$, below the condensation threshold $\rho_c\simeq4.06$~\cite{Ge2026}. We denote the density before preparation by $\rho_i$. For preparations starting from a normal state, we adjust $\rho_i$ to obtain the prescribed final density. For protocols starting from a common input, the initial state is fixed and the net charge transfer is zero. In both cases, the final equilibrium is the same, while the scalar and electric profiles at release can differ.

\subsection{Excess free energy and horizon dissipation}
\label{sec:distance}

The comparison of relaxation can depend on the observable used to measure the approach to equilibrium~\cite{VuHayakawa2025}. Previous studies have used quantum complexity~\cite{Aditya2026}, subsystem relative entropy~\cite{AresRylandsCalabrese2025}, entanglement asymmetry~\cite{BeniniGodetSingh2025}, and quantum fidelity~\cite{YamashikaCalabreseAres2025Bosons}.

After the source is removed, the system relaxes at fixed temperature and charge density $\rho_f$. We use the excess probe free energy to measure its departure from the common normal state~\cite{Ge2026}. The generalized free energy follows from the conserved energy current on the fixed background~\cite{Tian2023}. For homogeneous fields with vanishing scalar source, its density per unit boundary area is
\begin{equation*}
 F_{\mathrm{probe}}(t)=\int_0^1\!\dd z\,
 \left[\frac12(A_t')^2+f|\psi'|^2+z|\psi|^2\right].
\end{equation*}
Evaluating this expression on the normal state in Eq.~\eqref{eq:normal} gives $F_{\mathrm{eq}}=\rho_f^2/2$. We therefore define the excess free energy by
\begin{equation}
 D(t)=F_{\mathrm{probe}}(t)-F_{\mathrm{eq}}.
 \label{eq:distance}
\end{equation}
The excess free energy includes both scalar and electric contributions to the departure from equilibrium.

After the source is removed, no energy is supplied through the AdS boundary. The excess free energy decreases through the flux into the black hole horizon,
\begin{equation}
 \dot D=-\mathcal F_H(t)=-2|\partial_t\psi(t,1)|^2.
 \label{eq:flux}
\end{equation}
The system relaxes to the common normal state, so $D(\infty)=0$. Integrating the energy balance gives
\begin{equation}
 D(t)=\int_t^\infty\mathcal F_H(s)\dd s\ge0.
 \label{eq:futureflux}
\end{equation}
Thus $D(t)$ is nonnegative and nonincreasing during relaxation. It measures the excess free energy that remains to dissipate through the horizon.

\section{Nonlinear slow response and cancellation}
\label{sec:cancellation}

\subsection{Cubic Maxwell feedback}

After source removal, the charge density is conserved, but the scalar and Maxwell fields remain coupled. The radial Maxwell constraint in Eq.~\eqref{eq:gauss} determines how the scalar field modifies the electric potential during relaxation. We write $A_t=\rho(1-z)+a_\psi$, where $a_\psi'(0)=0$ and $a_\psi(1)=0$. The constraint gives~\cite{LiTianZhang2013,Ge2026}
\begin{equation}
 a_\psi(z)=\int_0^1\mathcal G(z,s)j_\psi(s)\dd s,
 \label{eq:green}
\end{equation}
with $\mathcal G(z,s)=\max(z,s)-1$ and $j_\psi=\ii(\bar\psi\psi'-\psi\bar\psi')$. The radial phase current of the scalar field produces an electric correction that is quadratic in the scalar amplitude. This correction then acts back on the scalar field during relaxation.

To study how this feedback affects the relaxation modes, we introduce $w=\psi'$. The scalar equation is first order in time for this variable. Since $\psi(t,0)=0$, radial integration recovers the field as $\psi=\cI w$, with $(\cI w)(z)=\int_0^z w(s)\dd s$. Substituting the electric potential from Eq.~\eqref{eq:green} into Eq.~\eqref{eq:scalar} gives
\begin{equation}
 \partial_t w=\cB_\rho w+\cN_3(w),\label{eq:reduced}
\end{equation}
with
\begin{align}
 \cB_\rho&=\frac f2\partial_z+\frac{f'}2+\ii\rho(1-z)
            -\frac{\ii\rho+z}{2}\cI,\label{eq:generator}\\
 \cN_3(w)&=\ii a_w w+\frac{\ii}{2}a_w'\cI w.\label{eq:cubic}
\end{align}
The operator $\cB_\rho$ describes linear evolution in the normal state electric potential. The correction $a_w=a_{\cI w}$ is determined by the scalar profile. Its feedback on the scalar field gives the cubic term $\cN_3$. The ingoing metric has $g^{tt}=0$, so the scalar equation contains no separate $A_t^2\Phi$ term. Higher responses arise by iterating the cubic feedback. This coupling can change the amplitudes of the relaxation modes during evolution.

\subsection{Nonlinear slow amplitude}

The operator $\cB_\rho$ determines the linear relaxation modes of the normal state. The Maxwell current is quadratic in the scalar perturbation, so the linear scalar field evolves on $A_t=\rho(1-z)$ in the homogeneous sector at fixed density. A scalar mode has the form $u_n(z)e^{\lambda_nt}$. Its radial equation is
\begin{equation}
 fu_n''+[f'+2\ii\rho(1-z)-2\lambda_n]u_n'
 -(z+\ii\rho)u_n=0.\label{eq:qnm}
\end{equation}
The mode is regular at the future horizon and satisfies $u_n(0)=0$ in the absence of a source. These conditions determine the retarded poles~\cite{SonStarinets2002,AmadoKaminskiLandsteiner2009}. We normalize $u_n'(0)=1$ and write $\lambda_n=-\gamma_n-\ii\Omega_n$ with $\gamma_n>0$.

At the final density $\rho=3$, the two slow modes used in the pulse construction are
\begin{align}
 \lambda_1&\simeq-0.177-0.373\ii,\label{eq:frequencies}\\
 \lambda_2&\simeq-1.77-1.59\ii.\nonumber
\end{align}
The second mode decays about ten times faster than the first. Suppressing the leading mode can therefore reveal a much faster stage of relaxation. Their different complex frequencies determine the amplitude and phase accumulated between boundary pulses. These frequencies enter the cancellation conditions below. Preparations with the same final normal state share these mode frequencies. The numerical calculation is described in Appendix~\ref{app:numerics}. The linear counterpart of slow mode suppression has been used in other settings~\cite{Klich2019,Carollo2021,Kochsiek2022,Ivander2023}. A related mechanism occurs in closed chaotic quantum systems, where reduced overlap with the dominant resonance accelerates subsystem equilibration~\cite{YamashikaHamazaki2026RP}.

To extract mode amplitudes from the evolving radial field, we use projections onto left modes. On horizon regular slices, the QNMs are eigenfunctions of the linear evolution operator~\cite{Warnick2015}. In the $w$ variable, $r_n=u_n'$ is a right eigenfunction of $\cB_\rho$. The corresponding left functional $\ell_n$ is normalized by $\ell_nr_n=1$, and $c_n(t)=\ell_nw(t)$ is the instantaneous modal projection. For the first mode, the nonlinear evolution gives $\dot c_1=\lambda_1c_1+\ell_1\cN_3(w)$. The first term describes linear evolution, while the second represents nonlinear excitation of the slow mode. The slow projection at release therefore does not determine the full late time response.

We set $t=0$ at source removal and write $w_0=w(0)$. For sufficiently small decaying solutions, the complete slow amplitude is $\cC_1(w_0)=\lim_{t\to\infty}e^{-\lambda_1t}c_1(t)$. It is the coefficient of the slow exponential at late times. Integrating the projected evolution gives
\begin{equation}
 \cC_1(w_0)=\ell_1w_0+
 \int_0^\infty e^{-\lambda_1s}\ell_1\cN_3(w(s))\dd s,
 \label{eq:asymptotic}
\end{equation}
The first term is the slow component present at release. The integral accounts for the slow response generated by nonlinear feedback afterward. The cancellation condition is
\begin{equation}
 \cC_1(w_0)=0.\label{eq:strongcondition}
\end{equation}
The asymptotic amplitude determines whether the slow exponential survives at late times. Its vanishing does not require the slow projection to vanish at release. The initial projection can instead compensate for the slow response generated by subsequent nonlinear evolution.
\subsection{Slow tail regeneration and compensation}

A simple example shows how nonlinear feedback regenerates the slow response. Start from a pure fast mode $w_0=b_0r_2$ with no initial slow projection. Its leading evolution is $w(t)\simeq b_0r_2e^{\lambda_2t}$. The cubic feedback contains two fields and one complex conjugate field. The resulting forcing has exponent $\mu_3=2\lambda_2+\bar\lambda_2$ and coupling $\kappa_2=\ell_1\cN_3(r_2)$. Integration gives
\begin{equation}
 c_1(t)=b_0|b_0|^2\kappa_2
 \frac{e^{\mu_3t}-e^{\lambda_1t}}{\mu_3-\lambda_1}
 +\ord{|b_0|^5}.\label{eq:regeneration}
\end{equation}
The nonlinear forcing from the fast mode decays at the rate $3\gamma_2$. The slow response excited by this forcing at early times continues to decay at the much smaller rate $\gamma_1$. The forcing can therefore decay quickly while leaving a slow response behind.

Nonlinear excitation also appears in black hole ringdown through other mechanisms~\cite{MaYang2024,KheraMaYang2025,BourgEtAl2025,OkuzumiIokaSakagami2008,LingShahWong2025,Sberna2022}. Here the cubic feedback is supplied by the constrained Maxwell field on a fixed geometry. The initial slow component can be chosen as
\begin{equation}
 c_1(0)=b_0|b_0|^2
             \frac{\kappa_2}{\mu_3-\lambda_1}\label{eq:compensation}
\end{equation}
An initial slow component of the form in Eq.~\eqref{eq:compensation} compensates the response generated afterward through cubic order. A state with no slow projection at release can therefore develop a slow tail, while a suitable initial component can suppress its leading nonlinear contribution. Figure~\ref{fig:cubic} shows the resulting slow projection.

\par\addvspace{4pt}\noindent\begin{minipage}{\columnwidth}
 \setlength{\abovecaptionskip}{4pt}
 \centering
 \includegraphics[width=\columnwidth]{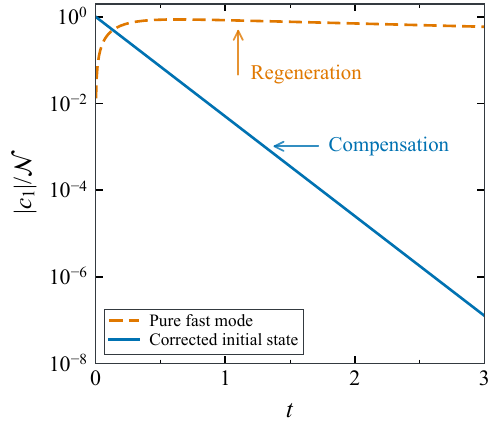}
 \captionof{figure}{Regeneration of a slow tail by a pure fast mode and its compensation by an initial slow component. The projection is normalized by $\mathcal N=|b_0|^3|\kappa_2/(\mu_3-\lambda_1)|$.}
 \label{fig:cubic}\label{fig:cubic:b}
\end{minipage}\par\addvspace{8pt}

The quantity to be compensated is the complete slow amplitude in Eq.~\eqref{eq:asymptotic}, rather than the slow projection at release. In the next section we construct boundary pulses that generate the required correction while accounting for the field evolution during preparation.

\section{Preparation with two pulses and the fast relaxation window}
\label{sec:preparation}

\subsection{Density history and the cubic pulse correction}

We use two boundary pulses to control the slow response. The first pulse excites the relaxation modes, while the second is chosen to cancel the slowest mode at linear order. Nonlinear feedback modifies this cancellation condition and requires a further correction to the second pulse. The source takes the form
\begin{equation}
 J(t)=\epsilon h_\sigma(t)+\alpha_2h_\sigma(t-\Delta).
 \label{eq:twopulse}
\end{equation}
The profile is $h_\sigma(t)=\sin^4(\pi t/\sigma)$ for $0<t<\sigma$ and vanishes outside this interval. The parameter $\epsilon$ is real. The complex coefficient $\alpha_2$ provides two independent controls of the first mode. The pulses do not overlap. The representative protocol has width $\sigma=0.5$, separation $\Delta=1$ and release time $t_0=\sigma+\Delta$.

During the drive, $\psi=J+\cI w$. The scalar evolution and the constrained electric response give
\begin{equation}
\begin{split}
 \partial_t w={}&\cB_\rho w+b_\rho J+\ii a_{w,J}w\\
 &+\frac{\ii}{2}a_{w,J}'(J+\cI w),
\end{split}
 \label{eq:forced}
\end{equation}
where $b_\rho=-(\ii\rho+z)/2$ and $a_{w,J}$ is evaluated on $J+\cI w$. This equation and the Ward identity evolve the density and the radial profile together.

Holographic pump probe studies have shown that the frequency content of a driving pulse controls the excitation of slow quasinormal modes~\cite{Bagrov2018Thermalization,Bagrov2018PumpProbe}. For a single pulse, the linear transfer to mode $n$ is
\begin{equation}
 H_n(\sigma)=(\ell_nb_{\rho_*})
       \int_0^\sigma e^{\lambda_n(\sigma-s)}h_\sigma(s)\dd s.
 \label{eq:mode-transfer}
\end{equation}
The factor $\ell_n b_{\rho_*}$ gives the coupling between the source and the mode. The integral propagates the response generated at each time to the end of the pulse and adds the resulting complex amplitudes. At the common release time, the response to the first pulse has evolved for an additional interval $\Delta$. The linear amplitude of mode $n$ is therefore $H_n(\sigma)[\epsilon e^{\lambda_n\Delta}+\alpha_2]$. Since $H_1\ne0$, the choice $\alpha_{2,\mathrm L}=\epsilon r_p$ with $r_p=-e^{\lambda_1\Delta}$ cancels the first mode. The remaining mode amplitudes are $\epsilon A_n^{(1)}$, with
\begin{equation}
 A_n^{(1)}=H_n(\sigma)
       [e^{\lambda_n\Delta}-e^{\lambda_1\Delta}].
 \label{eq:robust-mode-transfer}
\end{equation}
We adjust the initial density $\rho_i$ so that the drive ends at $\rho_f=3$.

\begin{figure*}[t]
 \begingroup
 \setlength{\abovecaptionskip}{4pt}
 \captionsetup[subfloat]{labelformat=empty,nearskip=0pt,farskip=0pt}
 \centering
 \smash{\makebox[0pt][l]{\subfloat[\label{fig:preparation-history:a}]{}\subfloat[\label{fig:preparation-history:b}]{}\subfloat[\label{fig:preparation-history:c}]{}}}
 \begin{tikzpicture}[inner sep=0pt,outer sep=0pt]
  \node[anchor=south west,inner sep=0pt,outer sep=0pt] (p) at (0,0) {\includegraphics[width=.96\textwidth,trim=0 29bp 0 0,clip]{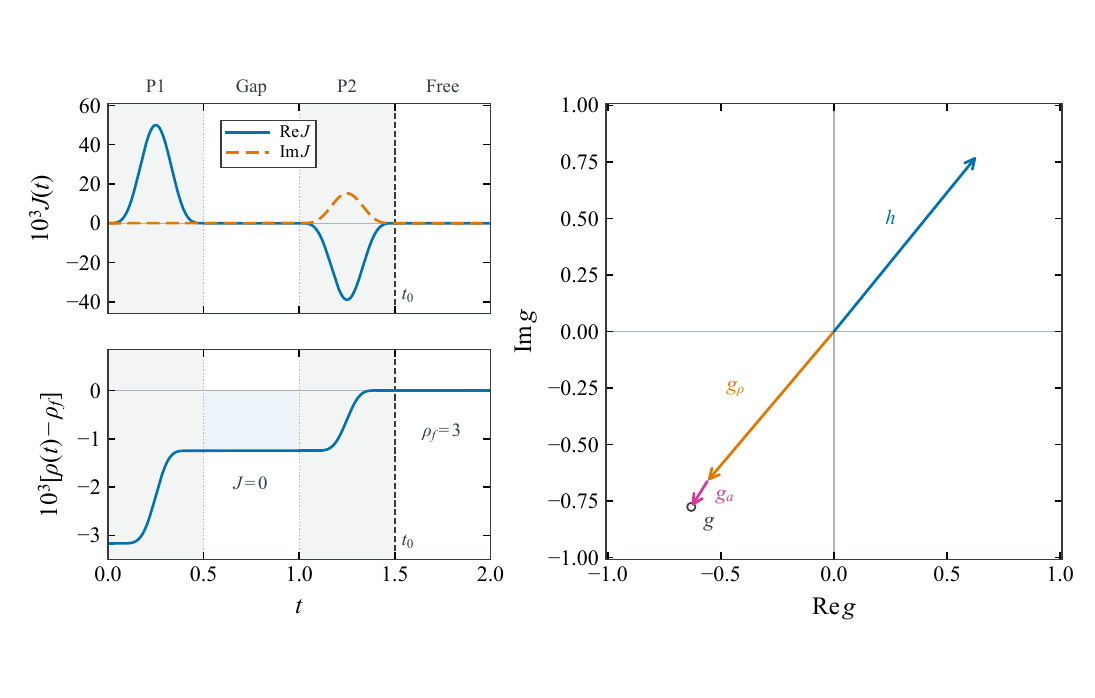}};
  \begin{scope}[shift={(p.south west)},x={($(p.south east)-(p.south west)$)},y={($(p.north west)-(p.south west)$)}]
   \node[anchor=south west,inner sep=0pt] at (.09863,0.83884) {\hypersetup{linkcolor=black}\normalsize\bfseries\subref{fig:preparation-history:a}};
   \node[anchor=south west,inner sep=0pt] at (.09863,0.45269) {\hypersetup{linkcolor=black}\normalsize\bfseries\subref{fig:preparation-history:b}};
   \node[anchor=south west,inner sep=0pt] at (.55342,0.83884) {\hypersetup{linkcolor=black}\normalsize\bfseries\subref{fig:preparation-history:c}};
  \end{scope}
 \end{tikzpicture}\par\nointerlineskip
 \caption{Preparation with two pulses at $(\sigma,\Delta)=(0.5,1)$, $\epsilon=0.05$ and $\rho_f=3$. The source follows the linear cancellation rule. \protect\subref{fig:preparation-history:a} Real and imaginary parts of the source. The labels P1 and P2 identify the two pulses. \protect\subref{fig:preparation-history:b} Density during the full nonlinear evolution. Shading marks the pulse intervals and the dashed line marks source removal. \protect\subref{fig:preparation-history:c} Cubic contributions in the complex plane. We use $g_\rho=G_\rho^{\rm prep}/|G|$, $g_a=(G_a^{\rm prep}+G_a^{\rm post})/|G|$, $g=G/|G|$ and $h=H_1\beta_3/|G|=-g$. The contributions $g_\rho$ and $g_a$ add head to tail. The compensation vector $h$ starts at the origin and points to $-g$. The terms are complex amplitudes.}
 \label{fig:preparation-history}
 \endgroup
\end{figure*}

Charge exchange and Maxwell feedback modify this linear rule. The leading scalar response is $O(\epsilon)$. The charge transfer and electric correction are $O(\epsilon^2)$, so their feedback first changes the scalar field at $O(\epsilon^3)$. We write $\alpha_2=\epsilon r_p+\epsilon^3\beta$ and expand
\begin{equation}
 \psi=\epsilon\psi_1+\epsilon^3\psi_3+O(\epsilon^5),\qquad
 \rho=3+\epsilon^2\rho_2+O(\epsilon^4).
 \label{eq:pulseexpansion}
\end{equation}
The leading field obeys $\dot w_1=\cB_{\rho_*}w_1+b_{\rho_*}J_1$, where $J_1=h_\sigma+r_ph_\sigma(t-\Delta)$ and $\psi_1=J_1+\cI w_1$. The Ward identity and the final density condition give
\begin{align}
 Q_2(t)&=2\Imag\!\left\{J_1^*[w_1(t,0)-\dot J_1+3\ii J_1]\right\},\nonumber\\
 \rho_2(t)&=-\int_t^{t_0}Q_2(s)\dd s,\qquad 0\le t\le t_0.
 \label{eq:preparation-density}
\end{align}
The condition $\rho_2(t_0)=0$ determines the initial density correction needed to reach the prescribed final density. During preparation, however, the charge density can still vary with time. This density history changes the scalar response even though the final density is fixed, as shown in Fig.~\ref{fig:preparation-history}.

The correction to the scalar field at cubic order obeys
\begin{equation}
 \dot w_3=\cB_{\rho_*}w_3+b_{\rho_*}\beta h_\sigma(t-\Delta)
                    +F_a+F_\rho,
 \label{eq:third-order-preparation}
\end{equation}
where the forcing separates into the radial electric response and the density history,
\begin{align}
 F_a&=\ii a_2w_1+\frac{\ii}{2}a_2'\psi_1,\nonumber\\
 F_\rho&=\ii\rho_2[(1-z)w_1-\psi_1/2].
 \label{eq:pulseforcing}
\end{align}
Here $a_2''=\ii(\bar\psi_1w_1-\psi_1\bar w_1)$ with $a_2'(0)=a_2(1)=0$. Even between pulses, the density excursion changes scalar propagation.

The response to the density history accumulates in the first mode during preparation. Its contribution at release is
\begin{equation}
 G_\rho^{\rm prep}=\int_0^{t_0}
      e^{-\lambda_1(t-t_0)}\ell_1F_\rho(t)\dd t.
 \label{eq:density-memory}
\end{equation}
Although the final density is fixed, its variation during the drive changes both the scalar evolution operator and the coupling to the boundary source. The propagation factor includes damping and phase accumulation for both density history and radial feedback. These effects accumulate during preparation and are encoded in the scalar profile at release.

The complete cubic amplitude also includes feedback after release. The cubic pulse correction contributes through the linear transfer $H_1\beta$, giving
\begin{equation}
 \begin{aligned}
 G&=G_a^{\rm prep}+G_\rho^{\rm prep}+G_a^{\rm post},\\
 \cC_1&=\epsilon^3(H_1\beta+G)+O(\epsilon^5).
 \end{aligned}
 \label{eq:pulse-cubic-amplitude}
\end{equation}
The coefficient $G_a^{\rm prep}$ comes from radial feedback during the drive, $G_\rho^{\rm prep}$ from the density history, and $G_a^{\rm post}$ from feedback after source removal. These are complex amplitudes. The preparation history is encoded in the released state rather than in an additional dynamical variable. At fixed density, the released scalar profile determines the subsequent evolution. The electric profile follows from the Maxwell constraint with the boundary conditions of Sec.~\ref{sec:model}. The correction that cancels their sum is
\begin{equation}
 \beta_3=-G/H_1.
 \label{eq:beta}
\end{equation}
This prescription uses the lower order fields and does not require nonlinear pulse tuning.

For the representative protocol at $\rho_f=3$, the density history contribution dominates. It is nearly in phase with the combined radial contribution $G_a=G_a^{\rm prep}+G_a^{\rm post}$, so the two interfere constructively. The density contribution has modulus $|G_\rho^{\rm prep}|/|G|\simeq0.87$ relative to the total. Appendix~\ref{app:perturbative-preparation} lists the representative complex coefficients.

The relative size of these contributions changes with the final density. Figure~\ref{fig:density-history-fraction} shows the modulus ratio for three protocols. For each curve, the pulse width and separation are fixed. At each final density, we recompute the slow mode and its projection and use the corresponding linear cancellation pulse. For $(\sigma,\Delta)=(0.5,1)$, the radial contribution relative to the density contribution decreases from about $0.4$ to $0.1$ as $\rho_f$ increases from $1$ to $4$. This decrease accounts for most of the increase in the plotted ratio. The shift toward aligned phases has a smaller effect and slightly offsets the increase.

\FloatBarrier
\par\addvspace{4pt}\noindent\begin{minipage}{\columnwidth}
 \setlength{\abovecaptionskip}{4pt}
 \centering
 \includegraphics[width=\columnwidth,trim=0 7bp 0 0,clip]{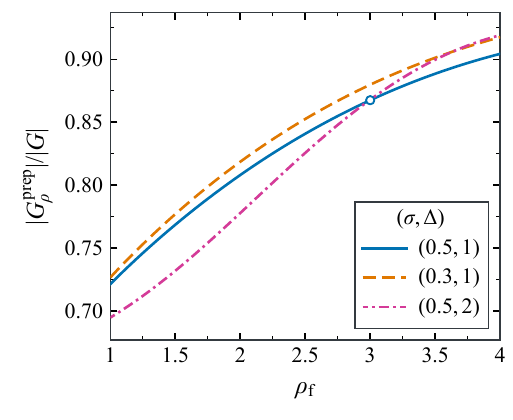}
 \captionof{figure}{Density dependence of $|G_\rho^{\rm prep}|/|G|$ for the three protocols with two pulses shown. The interval is $1\le\rho_f\le4$. The total $G$ includes the density history and radial feedback during preparation and after release. The open marker indicates $(\sigma,\Delta)=(0.5,1)$ at $\rho_f=3$. The plotted ratio compares moduli of complex amplitudes.}
 \label{fig:density-history-fraction}
\end{minipage}\par\addvspace{8pt}

\begin{figure*}[t]
 \begingroup
 \captionsetup[subfloat]{labelformat=empty,nearskip=0pt,farskip=0pt}
 \centering
 \plotpanelabove{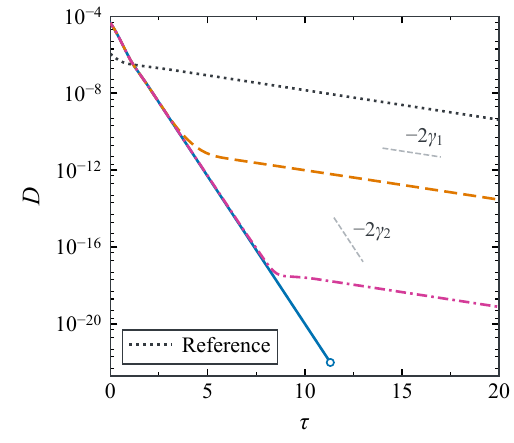}{fig:nlc-relaxation:a}{.49\textwidth}{.21482,.96508}\hfill
 \plotpanelabove{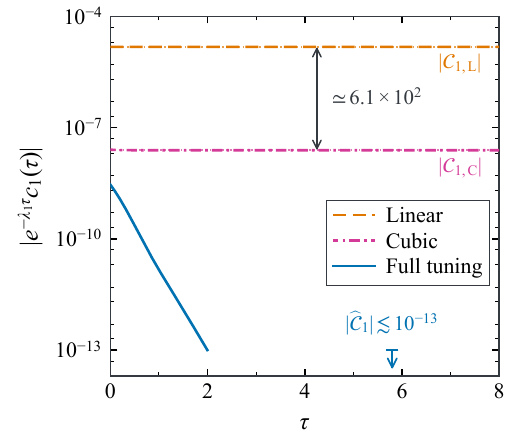}{fig:nlc-relaxation:b}{.49\textwidth}{.21482,.96508}\par\nointerlineskip
 \caption{Relaxation after preparation with two pulses at $\epsilon=0.05$ and $\rho_f=3$. Time is measured from source removal. \protect\subref{fig:nlc-relaxation:a} Excess free energy for the three preparations and the weak reference. The open marker indicates the plotting cutoff. \protect\subref{fig:nlc-relaxation:b} Magnitude of the slow projection with its linear time dependence factored out, $|e^{-\lambda_1\tau}c_1(\tau)|$. The arrow gives the numerical bound on $|\widehat{\cC}_1|$ for the fully tuned state. The vertical comparison gives the ratio of the linear and cubic slow amplitudes, approximately $6.1\times10^2$.}
 \label{fig:nlc-relaxation}
 \endgroup
\end{figure*}

\subsection{Relaxation reversal and the fast window}
\label{sec:boundary-signatures}

At a common final equilibrium, a prepared state can start with greater excess free energy and later overtake a reference that starts closer to equilibrium. We examine this reversal and how cubic compensation extends relaxation through faster channels.

We compare three preparations. The linear prescription cancels the slowest mode at linear order. The cubic prescription adds the correction predicted from the lower order equations, giving $\alpha_{2,\mathrm C}=\epsilon r_p+\epsilon^3\beta_3$. Full tuning adjusts the complex pulse coefficient and the initial density in the full nonlinear evolution to suppress the slow response while matching the final density. The cubic prescription suppresses the asymptotic slow tail but does not eliminate it. Full tuning seeks the condition in Eq.~\eqref{eq:strongcondition} within numerical resolution. The slow response is estimated from Eq.~\eqref{eq:finitetail}, with the remaining nonlinear contribution evaluated as described in Appendix~\ref{app:perturbative-preparation}. The leading residual orders for the first two prescriptions are
\begin{equation}
 \cC_{1,\mathrm L}=G\epsilon^3+O(\epsilon^5),\qquad
 \cC_{1,\mathrm C}=O(\epsilon^5).
 \label{eq:pulse-orders}
\end{equation}
All three protocols undergo the full nonlinear evolution and end at $\rho_f=3$. A weak single pulse provides the reference. Time is measured from source removal as $\tau=t-t_0$. Figure~\ref{fig:nlc-relaxation} displays the resulting distances and slow projections.

All three preparations overtake the weak reference. Linear pulse interference already produces the early crossing. Cubic compensation suppresses the residual slow amplitude and delays the subsequent dominance of the slow tail. It therefore extends the period governed by faster modes without changing their decay rates. For the representative protocol, cubic compensation reduces the slow amplitude by about a factor of $600$. For full tuning, the resulting bound is $|\widehat{\cC}_1|\lesssim10^{-13}$. The linear and cubic prescriptions eventually recover the slow slope $-2\gamma_1$.

For a single dominant mode, the boundary response and excess free energy behave as
\begin{equation}
 \langle\cO\rangle\sim A_ne^{\lambda_n\tau},\qquad
 D\sim K_n|A_n|^2e^{-2\gamma_n\tau}.
 \label{eq:obs-distance-ratio}
\end{equation}
where $K_n=|\lambda_nu_n(1)|^2/\gamma_n$. The field amplitude decays at rate $\gamma_n$, while the logarithmic decay rate of $D$ is $2\gamma_n$. When the weak reference is dominated by the first mode, its distance follows $D_r\sim K_1|A_{1,\mathrm{ref}}|^2e^{-2\gamma_1\tau}$.

The transition between fast and slow behavior also depends on interference. After the faster transients decay, write $\psi\simeq A_2u_2e^{\lambda_2\tau}+\cC_1u_1e^{\lambda_1\tau}$. The horizon balance gives
\begin{equation}
 D\simeq K_2|A_2|^2e^{-2\gamma_2\tau}
       +K_1|\cC_1|^2e^{-2\gamma_1\tau}+D_{12},
 \label{eq:two-mode-distance}
\end{equation}
with the interference contribution
\begin{equation}
 D_{12}=-4\Real\!\left[
 \frac{\mathfrak q_1\bar{\mathfrak q}_2e^{(\lambda_1+\bar\lambda_2)\tau}}
      {\lambda_1+\bar\lambda_2}\right],
 \label{eq:two-mode-interference}
\end{equation}
Here $\mathfrak q_1=\lambda_1\cC_1u_1(1)$ and $\mathfrak q_2=\lambda_2A_2u_2(1)$. The relative phase determines the sign of $D_{12}$. The faster decay of the prepared state allows it to overtake the weak reference. The interference term modifies the detailed crossing time. Related finite time crossings also occur in other relaxation settings~\cite{Chalas2024,Chatterjee2023}.

The faster mode can dominate after transients decay, but it also decays more rapidly. Even a small residual slow component can therefore take over at later times. Equating the two diagonal envelopes estimates the envelope takeover time as
\begin{equation}
 \tau_{\rm env}=\frac{1}{2(\gamma_2-\gamma_1)}
 \log\frac{K_2|A_2|^2}{K_1|\cC_1|^2}.
 \label{eq:distance-envelope}
\end{equation}
For the representative source, the envelope takeover time increases from about $4.2$ to $8.3$ under cubic compensation. The takeover time depends logarithmically on the ratio of the fast and slow amplitudes, so a large reduction in the slow amplitude leads to a finite delay in the return of slow mode dominance. These are envelope takeover times and differ from the crossing time of the two full distances. For nonzero leading coefficients, the slow contribution changes from $O(\epsilon^3)$ to $O(\epsilon^5)$, while the fast amplitude remains $O(\epsilon)$. This gives
\begin{equation}
 \begin{aligned}
 \tau_{\mathrm{env},\mathrm L}&=\frac{2}{\gamma_2-\gamma_1}\log\frac1\epsilon+O(1),\\
 \tau_{\mathrm{env},\mathrm C}&=\frac{4}{\gamma_2-\gamma_1}\log\frac1\epsilon+O(1).
 \end{aligned}
 \label{eq:window-amplitude-scaling}
\end{equation}
The coefficient of $\log(1/\epsilon)$ doubles as the slow tail is suppressed to higher order. The mode frequencies remain unchanged.

\subsection{Amplitude scaling}

\begin{figure}[t]
 \centering
 \includegraphics[width=\columnwidth]{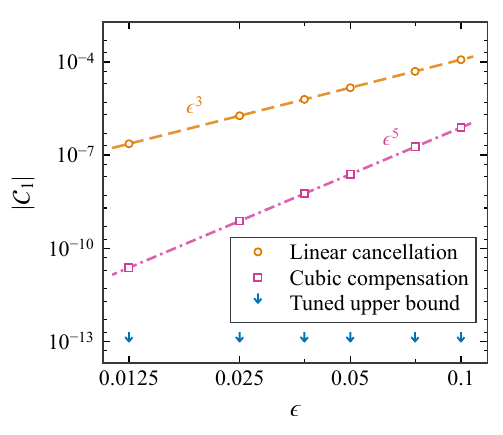}\par\nointerlineskip
 \caption{Residual slow amplitudes for the linear and cubic prescriptions at $\rho_f=3$. The guides show $|G|\epsilon^3$ and $|g_5|\epsilon^5$, where $g_5$ is estimated from the nonlinear evolution. The downward markers show numerical upper bounds for the fully tuned results.}
 \label{fig:nlc-scaling}
\end{figure}

We now test whether cubic compensation changes the residual slow amplitude from third to fifth order in the source strength. Figure~\ref{fig:nlc-scaling} shows six source amplitudes. Linear tuning leaves a cubic residual, and cubic compensation leaves a fifth order residual. The cubic guide uses a coefficient from lower order equations. The fifth order coefficient is estimated from full nonlinear evolution.

The higher order residual delays slow mode dominance and extends the fast window discussed in Sec.~\ref{sec:boundary-signatures}. The effective pulse correction obtained from full tuning approaches the cubic coefficient $\beta_3$. The difference between the linear and fully tuned fields also follows the predicted slow mode in its radial shape and complex phase. Appendix~\ref{app:perturbative-preparation} presents these comparisons in Figs.~\ref{fig:pulse-coefficients} and \ref{fig:radial-memory}.

We now turn to preparation that controls two slow channels.

\FloatBarrier

\section{Preparation with three pulses and interference between fast modes}
\label{sec:two-mode-preparation}

\begin{figure*}[t]
 \begingroup
 \captionsetup[subfloat]{labelformat=empty,nearskip=0pt,farskip=0pt}
 \centering
 \plotpanelabove{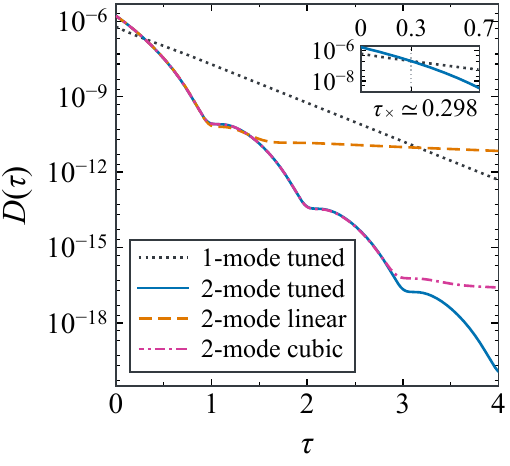}{fig:two-mode-preparation:a}{.32\textwidth}{.22750,.99395}\hfill
 \plotpanelabove{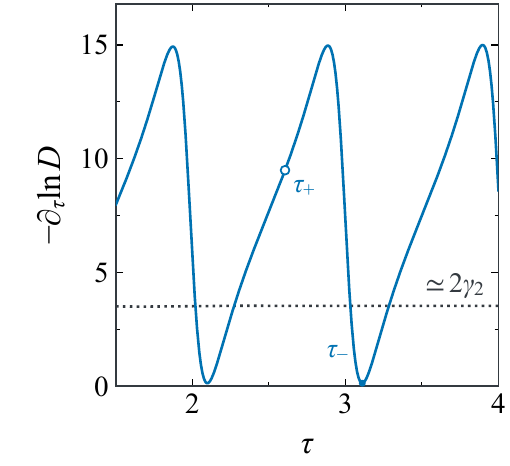}{fig:two-mode-preparation:b}{.32\textwidth}{.22750,.99395}\hfill
 \plotpanelabove{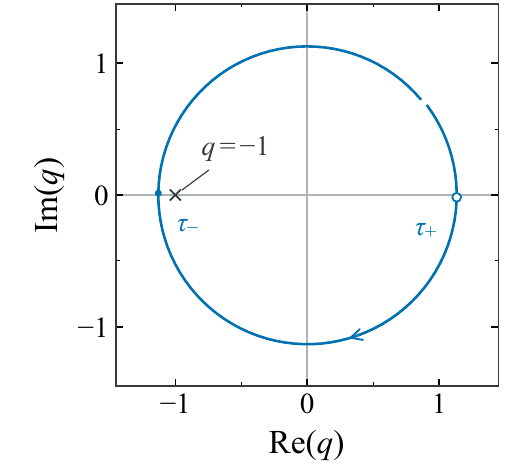}{fig:two-mode-preparation:c}{.32\textwidth}{.22750,.99395}\par\nointerlineskip
 \caption{Suppression of one and two slow channels at $\rho_f=3$, $\epsilon=0.05$, $\sigma=0.5$ and $\Delta=1$. The common release time is $t_0=2.5$. Linear and cubic refer to the two channel prescriptions. Tuned denotes full nonlinear tuning of one or two channels. \protect\subref{fig:two-mode-preparation:a} Excess free energy. The inset shows the crossing of the fully tuned states. \protect\subref{fig:two-mode-preparation:b} Instantaneous logarithmic decay rate $\mathcal F_H/D$. \protect\subref{fig:two-mode-preparation:c} Relative horizon phasor $q=h_4/h_3$ for $2.5\le\tau\le3.5$. Matching markers identify aligned phases near $\tau_+\simeq2.605$ and near cancellation at $\tau_-\simeq3.110$.}
 \label{fig:two-mode-preparation}
 \endgroup
\end{figure*}

Suppressing the first slow channel reveals the second. We therefore use three pulses to control both channels. The pulses start at $0$, $\Delta$ and $2\Delta$, with amplitudes $(\epsilon,\alpha_2,\alpha_3)$. The release time is $t_0=2\Delta+\sigma$. At release, their responses have evolved for an additional $2\Delta$, $\Delta$ and zero, respectively. For $E_n=e^{\lambda_n\Delta}$ and the transfer $H_n$ in Eq.~\eqref{eq:mode-transfer}, the linear amplitude of mode $n$ is
\begin{equation}
 A_n^{\rm lin}=H_n(\epsilon E_n^2+\alpha_2E_n+\alpha_3).
 \label{eq:three-pulse-polynomial}
\end{equation}
With the first pulse fixed, the complex coefficients $\alpha_2$ and $\alpha_3$ cancel the first two mode amplitudes. The two conditions give
\begin{equation}
 \alpha_{2,\mathrm L}=-\epsilon(E_1+E_2),\qquad
 \alpha_{3,L}=\epsilon E_1E_2.
 \label{eq:three-pulse-linear}
\end{equation}
These coefficients provide independent controls when $E_1\ne E_2$ and $H_1H_2\ne0$.

Nonlinear feedback generates slow responses during preparation and after release. A residual first slow mode also generates a contribution to the second modal projection that decays at $3\gamma_1<\gamma_2$. Cubic corrected coordinates distinguish this contribution from the response of the second mode. Their definition and extraction over the numerical window are given in Appendix~\ref{app:two-mode-extraction}.

Let $\boldsymbol G=(G_1,G_2)^T$ collect the cubic contributions to the two channels and let $\boldsymbol\beta_{\rm cl}=(\beta_2^{\rm cl},\beta_3^{\rm cl})^T$. For $\alpha_j=\alpha_{j,\mathrm L}+\epsilon^3\beta_j^{\rm cl}$ with pulse index $j=2,3$, the cubic residual is
\begin{equation}
 \boldsymbol C^{(3)}
 =\epsilon^3(\mathsf H\boldsymbol\beta_{\rm cl}+\boldsymbol G)
   +O(\epsilon^5),\qquad
 \mathsf H=
 \begin{pmatrix}H_1E_1&H_1\\H_2E_2&H_2\end{pmatrix}.
 \label{eq:two-mode-cubic}
\end{equation}
The two pulse corrections $\boldsymbol\beta_{\rm cl}=-\mathsf H^{-1}\boldsymbol G$ cancel both cubic responses when $\det\mathsf H=H_1H_2(E_1-E_2)\ne0$. The rows of $\mathsf H$ label the modes, while its columns describe corrections to the second and third pulses. The second pulse has an additional propagation factor $E_n$. The pulse corrections $\beta_j^{\rm cl}$ differ from the single coefficient $\beta_3$ in Sec.~\ref{sec:preparation}. The vector $\boldsymbol G$ is computed from the linear field and its density history. Each protocol adjusts the initial density so that $\rho_f=3$.

Across four source amplitudes, the cubic corrected coordinates $\widehat C_1$ and $\widehat C_2$ show the reduction from cubic to fifth order. Full tuning sets both coordinates at $\tau=3$ and the final density mismatch to zero. At $\epsilon=0.05$, the numerical bounds on their magnitudes are $10^{-14}$ and $4\times10^{-13}$, respectively. Table~\ref{tab:two-mode-residuals} gives the corresponding estimates.

We compare the two fully tuned preparations at the common release time $t_0=2\Delta+\sigma$. The one channel preparation is released earlier and evolves freely until this time. Figure~\ref{fig:two-mode-preparation}\subref{fig:two-mode-preparation:a} shows that the two channel preparation starts farther from equilibrium but overtakes the one channel preparation at $\tau\simeq0.30$. Suppressing the first slow channel exposes a distance decay rate near $2\gamma_2\simeq3.5$. When both channels are suppressed, faster modes appear with an average logarithmic decay rate of about $7.5$ over the displayed window. The instantaneous rate also oscillates, as shown in Fig.~\ref{fig:two-mode-preparation}\subref{fig:two-mode-preparation:b}.

At $\rho_f=3$, the fast modes have frequencies
\begin{align}
 \lambda_3&\simeq-3.7723+3.4555\ii,\label{eq:fast-frequencies}\\
 \lambda_4&\simeq-3.7801-2.7592\ii.\nonumber
\end{align}
Their horizon outputs interfere over the resolved window. Write $h_n=\lambda_nu_n(1)A_ne^{\lambda_n\tau}$ for the contribution of mode $n$ to $\partial_t\psi$ at the horizon. By Eq.~\eqref{eq:flux}, the coherent flux of the pair is
\begin{equation}
 \mathcal F_{34}=2|h_3+h_4|^2.
 \label{eq:fast-pair-reconstruction}
\end{equation}
Figure~\ref{fig:two-mode-preparation}\subref{fig:two-mode-preparation:c} shows the relative horizon phasor $q=h_4/h_3$. The close damping rates make its magnitude vary slowly, while the different frequencies rotate its phase. When $q$ approaches $-1$, the two horizon outputs nearly cancel, suppressing the instantaneous flux. Over $2.5\le\tau\le3.5$, the oscillation of the relaxation rate therefore reflects coherent interference between the two fast QNMs. The excess free energy continues to decrease monotonically.

\section{Charge neutral control from a common input}
\label{sec:common}

\begin{figure*}[t]
 \begingroup
 \captionsetup[subfloat]{labelformat=empty,nearskip=0pt,farskip=0pt}
 \captionsetup[subfloat]{farskip=0pt}
 \centering
 \plotpanelabove{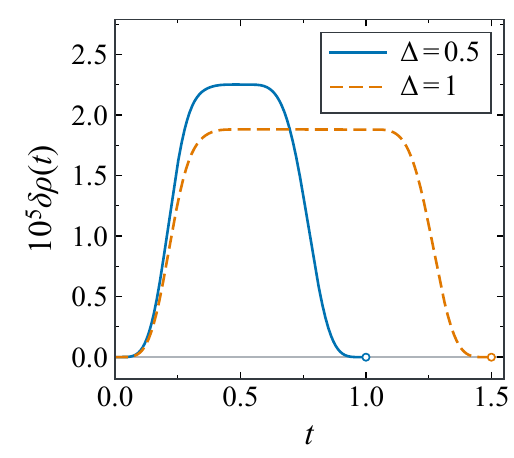}{fig:common:a}{.32\textwidth}{.22427,.95973}\hfill
 \plotpanelabove{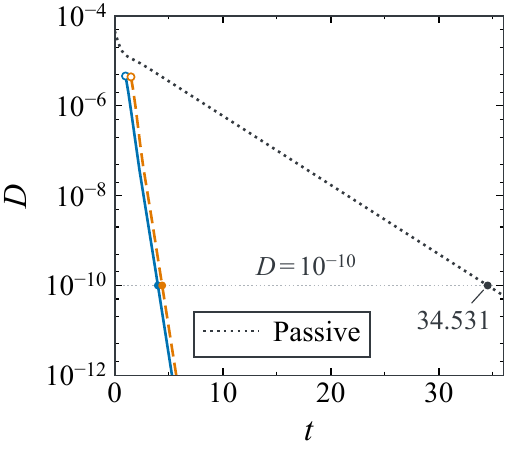}{fig:common:b}{.32\textwidth}{.22350,.96728}\hfill
 \plotpanelabove{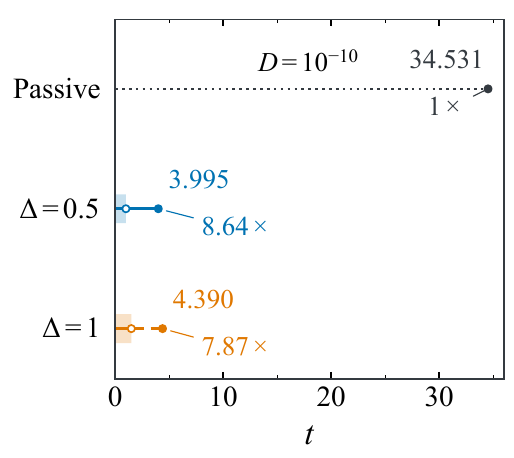}{fig:common:c}{.32\textwidth}{.22427,.95973}\par\nointerlineskip
 \caption{Charge neutral control from a common input with $\sigma=0.5$ and $\nu=3$. The common clock starts at release of the input. \protect\subref{fig:common:a} Charge change relative to the input. Open circles mark source removal. \protect\subref{fig:common:b} Distances after source removal, compared with passive evolution on the same clock. \protect\subref{fig:common:c} Total times to $D=10^{-10}$. Shading marks the full control duration. Open circles mark source removal and filled circles mark threshold arrival.}
 \label{fig:common}
 \endgroup
\end{figure*}

The preceding preparations adjust the initial density to reach the same final charge. We now fix the complete input state and require the control to return the transferred charge. The passive and controlled trajectories therefore start from the same radial scalar profile, constrained electric field and density. They approach the same normal equilibrium.

We take the release of the common input as $t=0$ and compare all subsequent trajectories on this clock, including the full duration of each control. The input is prepared by a single real pulse of width $0.5$ and amplitude about $0.032$. Its initial density is chosen so that the input has $\rho=3$ at release. Comparisons that include the complete control duration have also been studied in cavity quantum electrodynamics~\cite{Longhi2026Pontus}. Appendix~\ref{app:neutral-controls} gives the preparation and control parameters.

We use two pulses with a phase reset at their start times $t_1=0$ and $t_2=\Delta$. The source is
\begin{equation}
 J(t)=\sum_{j=1}^2\alpha_jh_\sigma(t-t_j)e^{\ii\nu(t-t_j)}.
 \label{eq:carriersource}
\end{equation}
On each pulse interval, the Ward identity gives
\begin{equation}
 \dot\rho=2\Imag(J^*\psi'(0))+2(\mu-\nu)|J|^2.
 \label{eq:carrierward}
\end{equation}
The leading normal background has $\mu=\rho=3$. Choosing $\nu=3$ therefore removes the second term at leading order. The source can still exchange charge with the system. A neutral protocol therefore requires the full Ward integral to vanish.

We adjust the two complex pulse amplitudes to suppress the slow response while returning the transferred charge to zero. At linear order, the slow response combines the freely evolving input with the responses to the two pulses. The leading charge constraint further restricts the pulse amplitudes. Together these leading order conditions can admit a continuous family of solutions, as described in Appendix~\ref{app:neutral-controls}. The protocols below satisfy the charge condition in the full nonlinear evolution.

Figure~\ref{fig:common} shows the charge return and compares the controlled and passive trajectories from the same input. At release, the controlled states are closer to equilibrium than the passive states at the same times, and their slow responses are also suppressed. The subsequent relaxation can therefore proceed through faster channels. Including the full control duration, the controlled trajectories reach $D=10^{-10}$ in about $4$ to $4.4$ time units, compared with about $35$ for passive relaxation. The total arrival time is reduced by approximately a factor of eight. This comparison follows the common input viewpoint of Ref.~\cite{NavaEgger2025}.

The source norm $\mathcal P_J=\int|J|^2\dd t=(35\sigma/128)(|\alpha_1|^2+|\alpha_2|^2)$ measures the strength of these nonoverlapping pulses. It is $9.367\times10^{-5}$ for $\Delta=0.5$ and $7.333\times10^{-5}$ for $\Delta=1$. The longer pulse separation reduces the source norm by $21.7\%$, but increases the total arrival time by about $0.4$. The norm is distinct from thermodynamic work and entropy production~\cite{Moroder2024,ChetriteKumarBechhoefer2021}. Related protocols also examine the speed of controlled relaxation~\cite{PelusoEggerNava2026}.

\FloatBarrier
\section{Conclusions and discussion}
\label{sec:discussion}

We have studied how finite boundary driving prepares rapidly relaxing states in a holographic system. At the same final charge density, suitable drives suppress the slowest relaxation channel and extend the fast relaxation window. In the common input protocols considered here, the controlled trajectories also reach a fixed threshold sooner when the full control duration is included. The nonlinear analysis shows why canceling the slow mode at linear order is not sufficient to eliminate its late tail. Charge exchange and radial Maxwell feedback during preparation affect the slow response, and nonlinear evolution continues to generate it after source removal. The quantity that controls the late slow tail is therefore the full asymptotic slow amplitude. It includes both the projection at release and the nonlinear response generated afterward. Preparing a rapidly relaxing state involves not only linear mode selection but also the preparation history and nonlinear coupling between modes.

This picture separates the relaxation spectrum from the history that determines mode excitation. Preparations ending at the same equilibrium share the same linear quasinormal frequencies, but their different driving histories leave different amplitudes in the relaxation channels. In this model, the lower order field equations determine the cubic contribution to the slow amplitude. A compensating pulse reduces the residual slow amplitude from $O(\epsilon^3)$ to $O(\epsilon^5)$. The compensation does not change the decay rate of the slow mode. Instead, it delays the time at which the residual slow tail takes over and allows faster channels to dominate for longer. Thus preparation can reshape relaxation over a finite time without changing the spectrum of the final state. The early distance crossing already appears under linear cancellation. The cubic correction mainly extends the subsequent fast window. Related studies have connected preparation history with anomalous relaxation in other systems~\cite{BaityJesi2019}. Nonlinear competition between quasinormal modes has also been studied after a density quench in a holographic superfluid~\cite{Ge2026}. By contrast, we calculate here the nonlinear slow response generated during and after finite boundary driving, and use a compensating pulse to suppress it.

When more than one slow channel is suppressed, coherent effects among the remaining modes become visible. Our preparation with three pulses suppresses two slow channels over the window studied. Faster modes then contribute to relaxation. Their different oscillation frequencies lead to interference between their coherent outputs at the horizon. As a result, the instantaneous logarithmic decay rate oscillates even though the excess free energy decreases monotonically. A single decay rate therefore need not describe the relaxation at every time. The relative phases of the modes also affect the instantaneous evolution.

The practical advantage of rapid relaxation also depends on the time needed for preparation. A smaller slow amplitude can extend the fast window, but the drive itself takes time and can exchange charge with the system. We therefore compared controlled and passive trajectories from the same complete input state under a condition of zero net charge transfer. For the protocols studied, the controlled trajectories reach the chosen free energy threshold sooner even when the full control duration is included. At release, the controls leave both a smaller excess free energy and a weaker slow response than passive evolution at the same time. Comparisons from a common input and studies of total protocol duration have also been considered in other systems~\cite{NavaEgger2025,Longhi2026Pontus}. These results suggest a practical comparison for experiments. In a dissipative system with controllable driving and several relaxation channels, one could vary the pulse amplitudes, phases and separations, then measure the slow tail, the fast relaxation window and the total time needed to reach a chosen threshold. Comparing conserved quantities and exchanges with the environment can help assess whether the trajectories approach the same final equilibrium. Experiments with trapped ions have demonstrated slow mode suppression~\cite{Zhang2025}, and coherent pulse spectroscopy has been used to control collective modes in a superconductor~\cite{Vaswani2021HiggsControl}. Such platforms could test slow tail regeneration and compensation in regimes where their effective dynamics contains nonlinear coupling between relaxation modes. Applying the comparison requires identifying the coupling of the experimental drive and the appropriate observable.

Our calculations use the probe limit to study charge exchange and nonlinear mode coupling on a fixed geometry. Beyond this approximation, the interaction between matter and geometry adds further dynamics to the preparation process. It remains to be studied whether slow tail regeneration and compensation persist with gravitational backreaction, and how the evolving geometry affects the preparation and its cost. Another direction concerns exceptional points, where mode coalescence and nonorthogonality can alter the excitation and coupling of relaxation channels~\cite{Zhou2023LEP,Zhang2026NonMarkovianEP,Chatterjee2024EP,LongEtAl2026EP}. These settings may require a different treatment of mode compensation. Studying them will help clarify how preparation controls relaxation in a broader class of strongly coupled systems.

\begin{acknowledgments}
We thank Dr.~Zehua Tian for helpful discussions.
This work was supported by the National Natural Science Foundation of China under
Grant Nos. 12305063, 12547143, and 12375047, and by the Hunan Provincial Natural Science
Foundation of China under Grant No. 2026JJ30135.
\end{acknowledgments}

\section*{Data availability}
The numerical solutions, figure data, and custom numerical and plotting scripts are available from the corresponding author upon reasonable request.

\appendix
\section{Nonlinear response and preparation details}
\label{app:numerics}

\begin{figure*}[t]
 \begingroup
 \captionsetup[subfloat]{labelformat=empty,nearskip=0pt,farskip=0pt}
 \centering
 \plotpanelabove{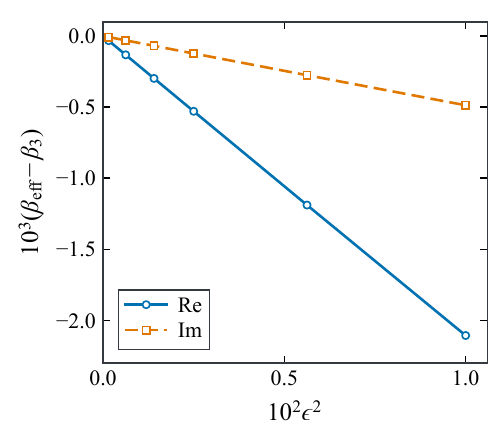}{fig:nlc-scaling:b}{.49\textwidth}{.20501,.952}\hfill
 \plotpanelabove{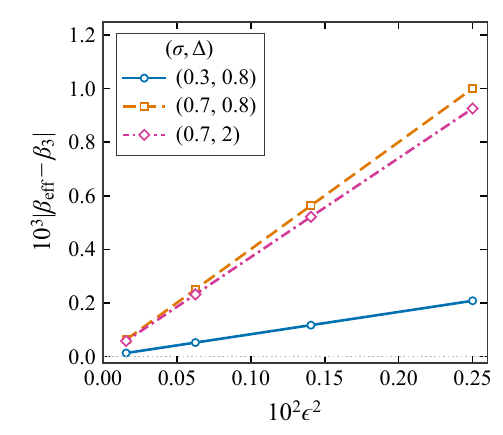}{fig:protocol-validation}{.49\textwidth}{.20501,.952}\par\nointerlineskip
 \caption{Pulse corrections at $\rho_f=3$. \protect\subref{fig:nlc-scaling:b} Real and imaginary parts of $\beta_{\rm eff}-\beta_3$ for the representative protocol. \protect\subref{fig:protocol-validation} Magnitude of the same offset for three additional protocols. The markers are full evolution results and the lines guide the eye.}
 \label{fig:pulse-coefficients}
 \endgroup
\end{figure*}

\subsection{Evolution and quasinormal modes}
\label{app:evolution}

We solve the scalar evolution equation and the radial Maxwell constraint in ingoing Eddington Finkelstein coordinates, with the boundary conditions specified in Sec.~\ref{sec:model}. The radial direction is discretized with $N=50$ Chebyshev points~\cite{Trefethen2000}. Boundary lifting supplies the source $J(t)$, and the Ward identity evolves the charge density through preparation and release. The scalar equation is imposed at the regular future horizon, where it directly supplies $\partial_t\psi(1)$.

To evaluate the excess free energy, we subtract the electric contribution of the common normal state in Eq.~\eqref{eq:distance}. Using the potential decomposition of Sec.~\ref{sec:cancellation} gives
\begin{equation}
 D=\int_0^1\!\dd z\,\left[-\rho_f a_\psi'
 +\frac12(a_\psi')^2+f|\psi'|^2+z|\psi|^2\right].
 \label{eq:distance-numerical}
\end{equation}
This expression is used to compute $D$ after source removal. The horizon flux provides a check of the energy balance.

The source free linear evolution matrix acts on scalar values at the nonboundary nodes. Its right eigenvectors give the radial modes, and its left eigenvectors give their projections. We normalize each right mode by $u_n'(0)=1$ and the corresponding discrete left row by $\ell_n^{\rm disc}u_n=1$. To evaluate $\ell_n w$, we reconstruct $\psi=\mathcal I w$ and apply $\ell_n^{\rm disc}$. Four modes are retained for the two channel diagnostic.

The radial equation~\eqref{eq:qnm} also admits a regular horizon series~\cite{HorowitzHubeny2000}. Write $u=\sum_{n=0}^M c_n(1-z)^n$ with $c_0=1$ and $c_{-1}=0$. The series coefficients satisfy
\begin{equation}
 c_{n+1}=\frac{[3n(n+1)+1+\ii\rho(2n+1)]c_n-n^2c_{n-1}}
 {(n+1)[3(n+1)+2\lambda]}.
 \label{eq:recurrence}
\end{equation}
The source free boundary condition is $\sum_{n=0}^M c_n=0$. To remove poles associated with the horizon normalization, we multiply this boundary series by $1/\Gamma(1+2\lambda/3)$. The two slow poles of the collocation spectrum are checked with this series. The fast modes used in the reconstruction come from collocation and are checked by radial shooting. The radial modes are rescaled to $u_n'(0)=1$.

\begin{table}[b]
 \centering
 \caption{Cubic pulse corrections and estimates of fifth order residual coefficients for four representative protocols at $\rho_f=3$. All four are tested by nonlinear evolution. The estimates of $g_5$ use the free intercept of $\cC_{1,\mathrm C}/\epsilon^5$ from $\epsilon=0.025,0.0375,0.05$.}
 \label{tab:protocol-validation}
 \begingroup
 \setlength{\tabcolsep}{0.5em}
 \renewcommand{\arraystretch}{1.15}
 \begin{tabular}{crr}
 \toprule
 $(\sigma,\Delta)$ & $\beta_3$ & $g_5$\\
 \midrule
 $(0.3,0.8)$ & $0.20889+0.01428\ii$ & $0.00815+0.01637\ii$\\
 $(0.5,1)$ & $0.33388+0.01136\ii$ & $0.03672+0.06908\ii$\\
 $(0.7,0.8)$ & $0.46588+0.03026\ii$ & $0.09930+0.17153\ii$\\
 $(0.7,2)$ & $0.39426-0.08078\ii$ & $0.11234+0.14500\ii$\\
 \bottomrule
 \end{tabular}
 \endgroup
\end{table}

\subsection{Nonlinear mode amplitudes and pulse corrections}
\label{app:perturbative-preparation}
\label{app:protocol-validation}
\label{app:two-mode-extraction}

The cubic pulse correction follows from the linear driven field. This field determines the quadratic charge history through the Ward identity and the quadratic Maxwell response. Both enter the cubic forcing of Eq.~\eqref{eq:pulseforcing}. We choose the initial quadratic density so that $\rho_2(t_0)=0$. Projection during preparation and after release gives $G=G_a^{\rm prep}+G_\rho^{\rm prep}+G_a^{\rm post}$ in Eq.~\eqref{eq:pulse-cubic-amplitude}. The compensating pulse is then $\beta_3=-G/H_1$. For the representative protocol $(\sigma,\Delta)=(0.5,1)$ at $\rho_f=3$, the predicted correction is $\beta_3\simeq0.33388+0.01136\ii$. The rounded complex contributions are $10^3G_\rho^{\rm prep}\simeq-67.19-79.47\ii$, $10^3G_a^{\rm prep}\simeq-8.54-13.56\ii$, and $10^3G_a^{\rm post}\simeq0.017-0.016\ii$. Their sum is $10^3G\simeq-75.71-93.05\ii$. The density contribution is larger than the radial contribution during preparation, and their phases are nearly aligned. The mode is normalized by $u_1'(0)=1$, and the time after release is $\tau=t-t_0$. The postrelease forcing is integrated to $\tau=8$ for the representative calculation and to $\tau=10$ for the protocol predictions.

At a finite extraction time $\tau_c$, the evolution equation gives the full slow amplitude in Eq.~\eqref{eq:asymptotic} as
\begin{equation}
 \cC_1=e^{-\lambda_1\tau_c}\ell_1w(\tau_c)
 +\int_{\tau_c}^{\infty}e^{-\lambda_1s}\ell_1\cN_3(w(s))\dd s.
 \label{eq:finitetail}
\end{equation}
The first term contains the slow response accumulated up to the extraction time, while the integral accounts for the nonlinear response generated during the remaining evolution. To estimate the remaining integral, we expand the late field in the first two discrete modes and integrate the resulting cubic terms analytically to infinity. We denote the resulting estimate of $\cC_1$ by $\widehat{\cC}_1$. The left rows are those of the discrete evolution matrix. Estimates at $\tau_c=6,8,10$ are compared. Full two pulse tuning varies the real and imaginary parts of $\alpha_2$ and the initial density $\rho_i$ to solve $\Real\widehat{\cC}_1=\Imag\widehat{\cC}_1=0$ and $\rho_f=3$. The initial guess is the cubic prediction with matched density. We use $\tau_c=6$ for the representative protocol and $\tau_c=8$ for the three additional protocols. For the linear and cubic prescriptions, $\alpha_2$ is held fixed and only $\rho_i$ is adjusted to match the final density.

The fully tuned pulse also provides a comparison between the predicted cubic correction and the correction required by the complete nonlinear evolution. Subtracting the linear amplitude and dividing by $\epsilon^3$ gives
\begin{equation}
 \beta_{\rm eff}(\epsilon)\equiv
 \frac{\alpha_{2,*}-\alpha_{2,\mathrm L}}{\epsilon^3}
 =\beta_3+\ord{\epsilon^2}.
 \label{eq:pulse-tuned-limit}
\end{equation}
The real and imaginary parts of $\beta_{\rm eff}$ approach the corresponding components of $\beta_3$ as the source amplitude decreases. Figure~\ref{fig:pulse-coefficients} shows these component differences for the representative protocol and $|\beta_{\rm eff}-\beta_3|$ for three additional protocols. Table~\ref{tab:protocol-validation} lists the cubic corrections and residual estimates for the four protocols tested by nonlinear evolution.

The radial field gives a complementary view of the slow response. Let $\delta\psi=\psi_{\mathrm L}-\psi_*$ be the difference between the linear and fully tuned preparations. Their leading fields agree, so the difference first appears at cubic order. After the faster transients decay, its leading slow component is $\delta\psi\simeq\epsilon^3G\,u_1(z)e^{\lambda_1\tau}$. Figure~\ref{fig:radial-memory} compares $e^{-\lambda_1\tau}\delta\psi/\epsilon^3$ with $Gu_1(z)$ at $\tau=6$. The field difference has the predicted radial profile and complex phase without an additional amplitude or phase fit.

\par\addvspace{4pt}\noindent\begin{minipage}{\columnwidth}
 \centering
 \includegraphics[width=\columnwidth]{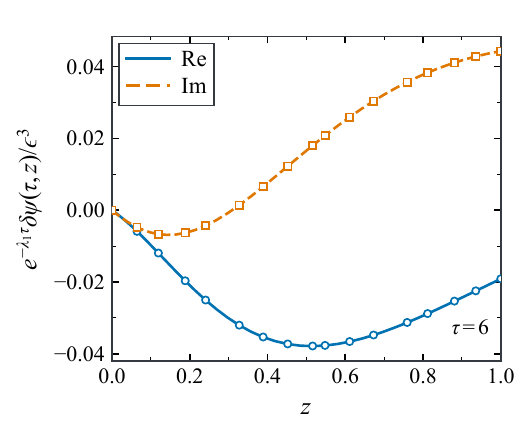}\par\nointerlineskip
 \captionof{figure}{Radial difference between the linear and fully tuned preparations at $\sigma=0.5$, $\Delta=1$, $\epsilon=0.05$ and $\rho_f=3$. At $\tau=6$, samples of $e^{-\lambda_1\tau}\delta\psi/\epsilon^3$ are compared with $Gu_1(z)$. The lines give the lower order prediction and the markers give the full evolution.}
 \label{fig:radial-memory}\label{fig:radial-memory:d}
\end{minipage}\par\addvspace{4pt}

For three pulses, the cubic forcing is projected onto both slow modes. The control matrix in Eq.~\eqref{eq:two-mode-cubic} then determines the two pulse corrections. A residual first mode can itself generate a contribution to the second projection. Its cubic self forcing decays at $3\gamma_1$, which is slower than the second mode because $3\gamma_1<\gamma_2$. Direct demodulation of the second projection therefore mixes this forced response with the second mode.

To separate the retained cubic contributions, we expand the field in four discrete modes and project the cubic Maxwell nonlinearity onto each of the first two left modes. The resulting terms have the form $T_{jabc}\bar c_a c_b c_c$. At linear order, their time dependence is governed by $\bar\lambda_a+\lambda_b+\lambda_c$. Applying an integrating factor to each term gives the frequency difference in the denominator below. The cubic correction separates the retained nonlinear forcing from the direct modal projections. These coordinates provide a finite order description of the two slow channels rather than an exact nonlinear decomposition. With $c_j=\ell_jw$ denoting the instantaneous modal projection, the corrected coordinates for $j=1,2$ are
\begin{equation}
 \begin{aligned}
 \widehat C_j={}&e^{-\lambda_j\tau}\Biggl[c_j+
 \sum_{a,b,c=1}^{4}
 \frac{T_{jabc}\overline{c_a}c_bc_c}
 {\lambda_j-\overline{\lambda_a}-\lambda_b-\lambda_c}\Biggr].
 \end{aligned}
 \label{eq:two-mode-normal-form}
\end{equation}
The denominators are nonzero for the retained terms. In the numerical extraction, each discrete projection is evaluated after subtracting the slower modes already extracted from the field. This sequential deflation reduces numerical leakage between modes and agrees with direct projection for biorthogonal modes in exact arithmetic. Omitted modes can also contribute, so $\widehat C_2$ is used as a finite window diagnostic over $2.5\le\tau\le3.5$. Full tuning varies the two complex pulse amplitudes and $\rho_i$ to set both corrected coordinates at $\tau=3$ and the final density mismatch to zero. Table~\ref{tab:two-mode-residuals} reports the corresponding estimates. The reconstruction in Fig.~\ref{fig:two-mode-preparation} uses the saved fast modes at $\tau=3$, with frequencies given in Eq.~\eqref{eq:fast-frequencies}.

\par\smallskip\noindent\begin{minipage}{\columnwidth}
 \centering
 \captionof{table}{Residual amplitudes of the two slow channels for linear cancellation, cubic compensation and full tuning at $\rho_f=3$. The corrected coordinates are evaluated at $\tau=3$. Full tuning is shown for $\epsilon=0.05$, with bounds set by numerical resolution. The second coordinate is a diagnostic within a finite time window.}
 \label{tab:two-mode-residuals}
 \begingroup
 \setlength{\tabcolsep}{0.5em}
 \renewcommand{\arraystretch}{1.15}
 \begin{tabular*}{211.72pt}{c@{\extracolsep{\fill}}lrr}
 \toprule
 $\epsilon$ & Design & $|\widehat C_1|$ & $|\widehat C_2|$\\
 \midrule
 $0.025$ & Linear & $1.70\!\times\!10^{-6}$ & $2.05\!\times\!10^{-7}$\\
 & Cubic & $8.11\!\times\!10^{-10}$ & $1.91\!\times\!10^{-10}$\\
 \addlinespace[3pt]
 $0.05$ & Linear & $1.36\!\times\!10^{-5}$ & $1.64\!\times\!10^{-6}$\\
 & Cubic & $2.59\!\times\!10^{-8}$ & $6.09\!\times\!10^{-9}$\\
 & Full & $\lesssim10^{-14}$ & $\lesssim4\!\times\!10^{-13}$\\
 \addlinespace[3pt]
 $0.075$ & Linear & $4.57\!\times\!10^{-5}$ & $5.53\!\times\!10^{-6}$\\
 & Cubic & $1.96\!\times\!10^{-7}$ & $4.61\!\times\!10^{-8}$\\
 \addlinespace[3pt]
 $0.1$ & Linear & $1.08\!\times\!10^{-4}$ & $1.31\!\times\!10^{-5}$\\
 & Cubic & $8.26\!\times\!10^{-7}$ & $1.94\!\times\!10^{-7}$\\
 \bottomrule
 \end{tabular*}
 \endgroup
\end{minipage}\par

\Needspace{15\baselineskip}
\subsection{Charge-neutral protocols}
\label{app:neutral-controls}

For charge neutral preparation from a common input, the pulse amplitudes must satisfy both slow mode cancellation and vanishing net charge transfer. Let $c_p$ be the slow projection of the freely evolving input and $s_j$ the transfer from pulse $j$. At the common release time, the linear slow amplitude is $\cC_1^{\rm lin}=c_p+s_1\alpha_1+s_2\alpha_2$. When $s_2\ne0$, the linear cancellation condition eliminates $\alpha_2$. The leading charge change then takes the form
\begin{equation}
 \Delta\rho^{(2)}=A_{\rm N}|\zeta|^2
                  +2\Real(\bar\zeta B_{\rm N})+C_{\rm N},
 \label{eq:chargeconic}
\end{equation}
where $\zeta=\alpha_1$ and $A_{\rm N}$ and $C_{\rm N}$ are real. For $A_{\rm N}\ne0$, completing the square gives a circle in the complex $\zeta$ plane when $|B_{\rm N}|^2-A_{\rm N}C_{\rm N}>0$. Equality gives a single point and a negative value gives no solution. These are leading order conditions. The protocols in Sec.~\ref{sec:common} satisfy charge neutrality in the full evolution.

The representative protocols use $(\sigma,\Delta,\epsilon)=(0.5,1,0.05)$ and $\rho_f=3$. The weak reference is a pulse of width $0.5$ and amplitude $0.005$. The common input comes from a real pulse of width $0.5$ and amplitude approximately $0.032$, with $\rho_i\simeq2.9992124$. All trajectories share its complete state, and the clock starts at its release. The charge neutral controls use $(\sigma,\nu)=(0.5,3)$ with $\Delta=0.5$ or $1$.

For the representative two pulse protocol, full tuning gives $\alpha_2\simeq-0.0389680+0.0152494\ii$ and $\rho_i\simeq2.9968335$. For the three pulse protocol, the tuned values are $\alpha_2\simeq-0.0388251+0.0237914\ii$, $\alpha_3\simeq-0.0027130-0.0066191\ii$ and $\rho_i\simeq2.9965932$. For the common input with $\Delta=0.5$, the control amplitudes are $\alpha_1\simeq-0.0202404+0.0102928\ii$ and $\alpha_2\simeq0.0045830+0.0121854\ii$. For $\Delta=1$, they are $\alpha_1\simeq-0.0170944+0.0103987\ii$ and $\alpha_2\simeq0.0037009+0.0110595\ii$. These parameters are rounded for display. The reported bounds are obtained with the unrounded tuned values.

We use $N=50$ radial Chebyshev points and DOP853~\cite{Hairer1993} for the driven evolution. The relative tolerance is $3\times10^{-13}$ for two pulses and $4\times10^{-13}$ for three pulses, with a maximum step of $0.005$. The common input is evolved after source removal with fourth order Runge Kutta and step $0.0025$. Checks at $N=60$ and $N=64$ support the two pulse and three pulse bounds, respectively. A half step comparison supports the common input arrival times. These bounds reflect finite numerical resolution.

\addtolength{\textheight}{2\baselineskip}
\bibliography{S-MPB}
\end{document}